\documentclass[twocolumn,showpacs,preprintnumbers,english,amsmath,amssymb]{revtex4}

\usepackage{graphicx,here}
\usepackage{dcolumn}
\usepackage{bm}
\def\simg{\,\hbox{\kern.1em \lower.6ex \hbox{$\sim$} \kern-1.12em
          \raise.6ex \hbox{$>$} }}

\begin{document}

\title{Supershell structure in trapped dilute Fermi gases}
\date{November 11, 2005}
\author{Y. Yu}
\author{M. \"{O}gren}
\author{S. \AA berg}
\author{S. M. Reimann}
\affiliation{Division of Mathematical Physics, LTH, Lund University. P.O. Box
118, S-221 00 Lund, Sweden}
\author{M. Brack}
\affiliation{Institut f\"{u}r Theoretische Physik, Universit\"{a}t
Regensburg, D-93040 Regensburg, Germany}

\begin{abstract}
We show that a dilute harmonically trapped two-component gas of
fermionic atoms with a weak repulsive interaction has a pronounced
{\it super-shell} structure: the shell fillings due to the spherical
harmonic trapping potential are modulated by a beat mode. This
changes the ``magic numbers'' occurring between the beat nodes
by half a period. The length and amplitude of this beating mode
depend on the strength of the interaction. We give a simple
interpretation of the beat structure in terms of a semiclassical
trace formula for the symmetry breaking U(3) $\rightarrow$ SO(3).
\end{abstract}

\maketitle

In finite systems of fermions, quantum effects lead to bunching of energy 
levels resulting in shell structure. Well-known examples are the shell 
structures in atoms or nuclei \cite{bm}, determining their chemical 
properties and stability. In fact, shell effects and the so-called
``magic numbers'', corresponding to spherical shell closings, have been
discovered in a variety of other finite fermion systems. Metal clusters
in which the delocalized valence electrons are bound in the field of
the metallic ions \cite{mbrack}, or quantum dots in semiconductor
heterostructures \cite{steffi} are famous examples. More recent
experimental progress makes it possible to study yet another species of
finite quantal systems: atomic gases, often weakly interacting, confined
e.g.\ by an optical dipole trap \cite{grimm}.

In this paper, we show that a harmonically trapped gas of fermionic atoms 
interacting by a weak repulsive two-body force may exhibit super-shell 
structure: the shell oscillations of the spherical harmonic oscillator are 
modulated by a beat structure, whereby the positions of the magic numbers 
are shifted by half a period between successive beats. We can explain this 
surprising result semiclassically by the interference of diameter 
and circle orbits surviving the breaking of the U(3) symmetry of the 
harmonic oscillator by the leading anharmonicity term in the mean field.

Similar super-shell structure was predicted 
for metallic clusters \cite{cluster}, inspired by a 
semiclassical analysis of Balian and Bloch in terms of the periodic orbits 
in a spherical cavity \cite{bablo}, and observed experimentally \cite{klavs}. 
Analogous ideas could be applied to the description of shell structure in 
transport properties of quantum wires \cite{wire}. 

Let us consider a dilute gas of fer\-mionic atoms, confined by a
spherical harmonic potential modeling an external trap
\cite{grimm}, interacting through a repulsive zero-range two-body
potential. The many-body Hamiltonian is
\begin{equation}
H=\sum_{i=1}^{N}\left(\frac{\mathbf{{p}}_{i}^{2}}
{2m}+\frac{m}{2}\,\omega^{2}\mathbf{{r}}_{i}^{2}\right)
+\frac{4\pi\hbar^{2}a}{m}\sum_{i<j}\delta^{3}
(\mathbf{{r}}_{i}-\mathbf{{r}}_{j})\,,\nonumber
\end{equation}
where $a$ is the $s$-wave scattering length. The value and the sign
of $a$ can be varied either by changing the type of atoms or by
applying a magnetic field: Feshbach resonances \cite{feshbach}
allow to tune the scattering length from a large positive to a
large negative value. (Here, we focus on the repulsive case.)

Due to the Pauli principle the $\delta$ interaction only applies to
fermions of pairwise opposite spin. We consider a fully unpolarized
two-component system with two spin states, so that the total particle
density is composed of two different densities of equal magnitude,
$n({\bf r})=n^{\uparrow}({\bf r})+n^{\downarrow}({\bf r})=
2n^{\uparrow}({\bf r})$. In the weak-interaction regime, the
interaction energy density is given by $gn^{\uparrow}({\bf r})
n^{\downarrow}({\bf r})=gn^2({\bf r})\!/4$, where the coupling strength
parameter $g$ is introduced by
\begin{equation}
g=4\pi\hbar^{2}a/m.
\end{equation}
This leads to the single-particle Hartree-Fock equation
\begin{equation}
\left[-\frac{\hbar^{2}}{2m}\Delta+gn^{\uparrow}({\bf r})+V_{ho}({\bf r})\right]
\psi_{i}^{\downarrow}({\bf r})=\epsilon_{i}\psi_{i}^{\downarrow}({\bf r})\,,
\label{HFI}
\end{equation}
where $V_{ho}$ is the harmonic oscillator (HO) trap potential.

\begin{figure}[H]
\begin{center}\includegraphics[%
  width=7cm]{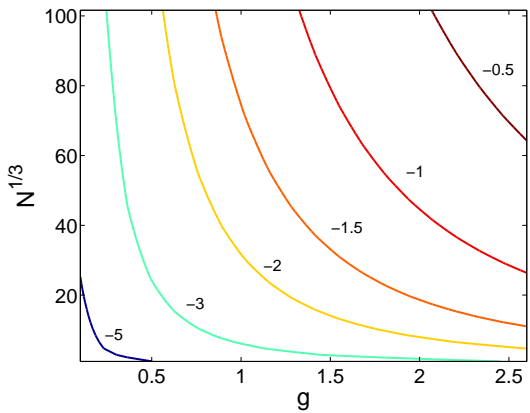}\end{center}
\vspace*{-0.5cm}
\caption{(Color online)Level curves of
$_{10}log\left(a^{3}n\left(0\right)\right)$ calculated
within the Thomas-Fermi approximation.  $g= 4 \pi \hbar^2 a/m$ 
($\hbar$=$\omega$=$m$=1).
} \label{figart1level}
\end{figure}

The diluteness condition necessary to treat the interaction as a two-body
process is that the interparticle spacing ${\bar n}^{-1/3}$ is much
larger than the range of the interaction and that ${\bar n}a^{3}\ll1$.
This dimensionless parameter also limits the life time of the
two-component atomic fermion gas due to dimer formation, which is a
three-body process \cite{petrov}. To guarantee that this condition is
fulfilled, we calculate the central density $n(0)$ in the Thomas-Fermi
approximation and plot level curves of $\log(a^3 n(0))$ in a $g$-$N$
landscape seen in Fig.\ \ref{figart1level}. The diluteness condition is
seen to be fulfilled for all considered combinations of particle numbers
and interaction strengths. This also outrules the possibility of phase 
separation discussed in \cite{phaseseparation}. 

Assuming spherical symmetry, Eq.(\ref{HFI}) reduces to its radial part
\begin{equation}
\left[-\frac{\hbar^{2}}{2m}\frac{1}{r^{2}}\frac{\partial}
{\partial r}\!\left(\!r^{2}\frac{\partial}{\partial r}\right)\!
+\frac{\hbar^{2}}{2m}\frac{l(l+1)}{r^{2}}+U(r)\right]
\psi_{i}^{\downarrow}=\epsilon_{i}\psi_{i}^{\downarrow},
\label{SSE}
\end{equation}
with $U(r)=gn^{\uparrow}(r)+\frac{1}{2}m\omega^{2}r^{2}$ being the 
effective mean-field potential. Each state has a $(2l+1)$-fold angular 
momentum degeneracy. We solve Eq.\ (\ref{SSE}) self-consistently on a  
grid. The interaction term is updated (with some weight factors) in each 
iteration according to $gn^{\uparrow}(r)=g\sum_i|\psi_{i}^{\uparrow}(r)|^2$.

\vspace*{-0.5cm}

After convergence is obtained, the ground-state energy of the
$N$-particle system is given by ($E_F$ = Fermi energy)
\begin{equation}
E_{tot}(g,N)=\sum_{\epsilon_i\leq E_F,\,\sigma=\uparrow,\downarrow}
                        \epsilon_{i}-g\int
                        n_{\uparrow}^2(r)\,d^3r\,.
\label{EHF}
\end{equation}
In general, the ground-state energy as a function of $N$ can be
written as the sum of a smooth average part and an oscillating part,
$E_{tot}=E_{av}+E_{osc}$. The oscillating part, referred to as the
shell-correction energy, or shell energy in short, reflects the
quantized level spectrum $\{\epsilon_i\}$. For a non-interacting
Fermi gas in a spherically symmetric $3D$ harmonic trap, the
leading-order term for the average energy is found in the Thomas-Fermi
approximation to be \cite{brack} $E_{av}^{ho}=(3N)^{4/3}\hbar\omega/4$.
For the repulsive interacting case, we find
$E_{av}\left(g>0\right)\propto N^{\alpha}$ with a larger exponent
$\alpha>4/3$. However, Eq.\
(\ref{HFI}) with an interaction term linear in the density is only
valid for moderate $g$ values and in practice we are close to
$\alpha=4/3$ (e.g., $\alpha\approx1.35$ for $g=2$). Contrary to the
non-interacting case, and also to self-saturating fermion systems
(such as nuclei and metal clusters) with a nearly constant particle
density, it is not possible here to obtain the smooth part of the
energy by a simple expansion in volume, surface and higher-order
terms. We therefore perform a numerical averaging of the energy
(\ref{EHF}) over the particle number $N$ in order to extract its
oscillating part.

In the non-interacting case ($g$=0) the shell energy $E_{osc}$
oscillates with a frequency $2\pi\,3^{1/3}\approx 9.06$ as a
function of $N^{1/3}$ and has a smoothly growing amplitude
$\propto N^{2/3}$. This follows from the exact trace formula \cite{brack}
for $E_{osc}$ of the 3D harmonic oscillator, whose leading-order
term is given by
\begin{equation}
E_{osc}^{ho} \simeq (3N)^{\frac23}\frac{\hbar\omega}{2\pi^2}
                    \sum_{k=1}^\infty \frac{(-1)^k}{k^2}\,
                    \cos\left(2\pi k\,(3N)^{\frac13}\right).
\label{tfho}
\end{equation}
Hereby $k$ is the repetition number of the primitive classical periodic
orbit of the system with action $S_0(E)=2\pi E/\omega$. The argument of
the cosine function in Eq.\ (\ref{tfho}) is simply $k$ times $S_0(E)/\hbar$,
taken at the Thomas-Fermi value of the Fermi energy
$E_F(N)=(3N)^{1/3}\hbar\omega$. The gross-shell structure is governed by
the lowest harmonic with $k=1$.

\begin{widetext}

\begin{figure}[htbp]
\begin{center}
\includegraphics[angle=0,width=14.cm]{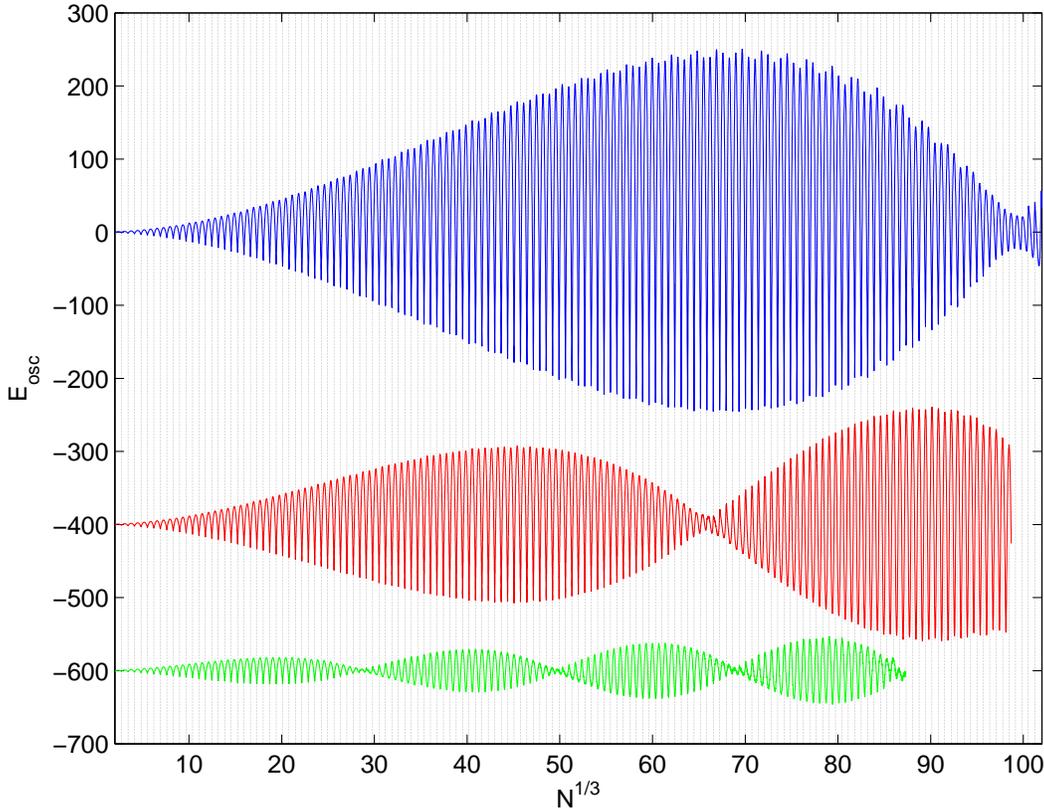}
\end{center}
\vspace*{-0.5cm}
\caption{ (Color online) The oscillating part of the ground state energy in units of
  $\hbar \omega$ as a
function of $N^{1/3}$ for $g= 0.2 $ (blue), 0.4 (red) and 2 (green).
The two lower curves are displaced by $400 \hbar \omega$ and $600
\hbar \omega$, respectively. The vertical dotted lines correspond to
the HO magic numbers $N_{mag}=M(M+1)(M+2)/3$ for $M=1,2,\dots$}
\label{figart}
\end{figure}

\end{widetext}

Switching on the interaction, this scenario changes. A beating
modulation of the rapid oscillations is found.
In Fig.\ \ref{figart} we show the shell energy versus $N^{1/3}$ for
three values of the interaction strength, $g$=0.2, 0.4 and 2. A beating
modulation of the amplitude of the shell energy, i.e., a {\it super-shell
structure}, is clearly seen to appear for all cases. At small particle
numbers and particularly for small $g$ values, the shell energy is that
of the non-interacting system, given by Eq.\ (\ref{tfho}). For larger
interaction strengths the super-shell structure is more clearly seen,
and several beating nodes appear for $g$=2. With increasing interaction
strength the amplitude of the shell energy oscillations becomes smaller.
For example, for particle numbers around $80^3 \approx 500 000$, the
amplitude of the shell energy is about 40 $\hbar\omega$, which is only
about $10^{-6}$ of the total ground-state energy.

\begin{figure}[H]
\begin{center}\includegraphics[%
  width=8.6cm,
  angle=0]{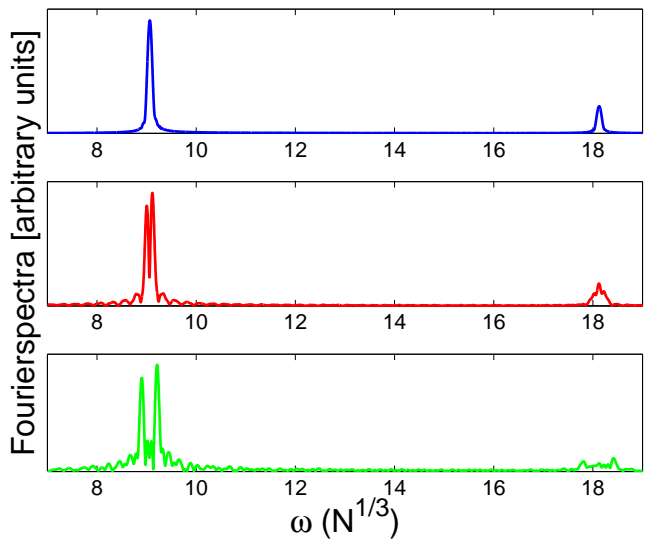}\end{center}
\vspace*{-0.3cm}
\caption{(Color online) Fourier spectra of the shell energy for $g=0.2$ (blue),
0.4 (red) and 2 (green). A peak splitting around $\omega= 9.06 $ is
resolved in the middle and bottom panels. The second harmonics ($k=2$)
are seen around $\omega= 18$. }
\label{figft}
\end{figure}

\begin{figure}[H]
\begin{center}\includegraphics[%
  width=7.7cm,
  angle=0]{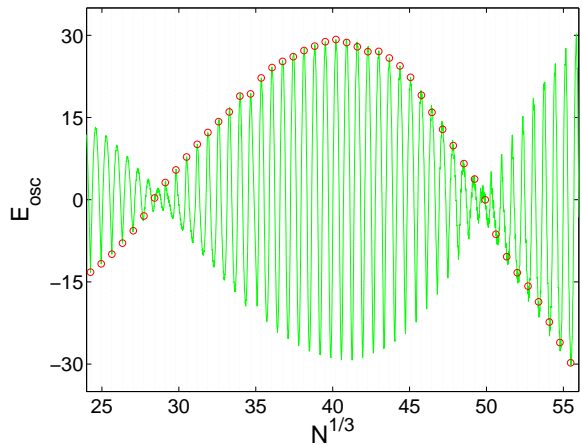}\end{center}
\vspace*{-0.7cm}
\caption{(Color online) An enlarged part of Fig.\ 2 for $g=2$. The circles mark the
harmonic oscillator magic numbers $N_{mag}$.}
\vspace*{-0.2cm}
\label{figart1g1}
\end{figure}

Through Fourier analysis of the calculated shell energy, two frequencies
are seen to smoothly appear with increasing $g$ value around the HO
frequency (9.06), see Fig.\ \ref{figft}. The exact values of the two
frequencies depend on the range of particle numbers included in the
analysis. The super-shell features appear when the contribution to the
effective potential from the interaction, $g n^{\uparrow}$, is
sufficiently large, i.e., at large values of $g$ and $N$. We also observe
that (almost) until the first super-node, i.e., $N^{1/3}\approx 28 $ in
Fig.\ \ref{figart1g1}, the magic numbers agree with the HO ones ($g=0$).
Between the first two super-nodes, i.e., $ 28\le N^{1/3}\le 49$ in Fig.\
\ref{figart1g1}, the magic numbers for the interacting system are
situated in the middle of two HO magic numbers, i.e., they appear at the
maxima of the fast shell oscillations. Then, after the second super-node
they roughly agree with the unperturbed HO ones again.

In the following we outline a semiclassical interpretation of these
features \cite{semicl}. The U(3) symmetry of the unperturbed HO system is
broken by the term $\delta U=gn^\uparrow$ in (\ref{HFI}), resulting in the
SO(3) symmetry of the interacting system. The shortest periodic orbits in
this system are the pendulating diameter orbits and the circular orbits
with a radius corresponding to the minimum of the effective potential in
(\ref{SSE}) including the centrifugal term. These two orbits lead to
the observed supershell beating. The above symmetry breaking has so far
not been discussed in the semiclassical literature. In a perturbative
approach \cite{crpert}, it can be accounted for by a group average of the
lowest-order action shift $\Delta S(o)$ brought about by the perturbation
of the system: $\langle e^{\frac{i}{\hbar}\,\Delta S(o)}\rangle_{o\in U(3)}$.
Hereby $o$ is an element of the group U(3) characterizing a member of the
unperturbed HO orbit family (ellipses or circles). For the average it is
sufficient to integrate over the 4-dimensional manifold $\mathbb{C}$P$^2$ 
\cite{bbz}, which for a perturbation $\delta U(r) = \varepsilon r^4$ can be 
done analytically \cite{semicl}.

\vspace*{-1.4cm}

\newpage

\vspace*{-1.0cm}

In the perturbative regime ($\varepsilon\ll 1$) we find the following
perturbed trace formula:
\begin{eqnarray}
E^{pert}_{osc}(N) = \frac{m^{\!2}\omega^4}{2\varepsilon\pi^3}\!
                    \sum_{k=1}^\infty \frac{(-1)^k}{k^3}\!
                    \left[\sin\!\left(\!\frac{kS_c}{\hbar}\!\right)\!
                    -\sin\!\left(\!\frac{kS_d}{\hbar}\!\right)\!\right]\!\!,
\label{tfpert}
\end{eqnarray}
where $kS_d$ and $kS_c$ are the classical actions of the diameter and
circle orbits, respectively. In the limit $\varepsilon\rightarrow0$,
their difference goes as $k(S_c-S_d)\rightarrow k\varepsilon\pi
E_F^2(N)/m^2\omega^5$, so that (\ref{tfpert}) tends to the pure HO limit 
(\ref{tfho}). Extracting $\varepsilon$ from a polynomial fit to the 
numerical potential $U(r)$ in (\ref{SSE}), one can qualitatively describe 
the beating of the shell energy $E_{osc}(N)$. With $\varepsilon$ of order 
$\sim5\times\!10^{-4}g$, Eq.\ (\ref{tfpert}) approximately reproduces the 
curves seen in Fig.\ \ref{figart} up to the beginning of the second 
supershell. It explains, in particular, also the phase change in the 
position of the magic numbers $N_{mag}$ shown in Fig.\ \ref{figart1g1}.

To cover larger values of $\varepsilon$ (and $N$), we have developed 
\cite{semicl} an analytical uniform trace formula for the potential 
$U(r)=m\omega^2r^2\!/2+\varepsilon r^4$, which contains the contributions 
of the 2-fold degenerate families of diameter and circular orbits to all
orders in $\varepsilon$. This is analogous to uniform trace formulae 
obtained earlier for U(1) \cite{toms} and U(2) symmetry breaking \cite{hhuni}.

The beat structure in $E_{osc}$ has some similarities with that found 
in nuclei \cite{bm} and metal clusters \cite{mbrack}. There are, however, 
two essential differences. 1. Those systems are self-saturating and have 
steep mean-field potentials that can be modeled by a spherical cavity 
\cite{bablo}. The present system, in contrast, has a mean field with much 
smoother walls that are dominated at large distances by the confining 
harmonic potential. 2. The super-shells in the cavity model come from the 
interference of the shortest periodic orbit families with three-fold
degeneracy, as is usual in spherical systems \cite{bertab,crli}. Here, however, 
the gross-shell structure comes from the diameter and circle orbits which 
are only 2-fold degenerate, whereas the fully 3-fold degenerate families of 
tori with rational ratios $\omega_r:\omega_\varphi=n:m$ of radial and angular
frequency only contribute to the finer quantum structures at higher energies. 
As described in \cite{semicl}, they bifurcate from the circle orbit with 
repetition numbers $k\geq 3$ and can be included in the trace formula using 
standard techniques \cite{bertab,crli,kaidel}.

As mentioned above, in a Fermi gas of atoms with repulsive interaction
($a>0$), atoms can be lost through three-body recombination events. Two
atoms with opposite spin form a molecule while the third takes up
energy. Having a low recombination rate, and thus a long life time
of the system, is desired.  Petrov  \cite{petrov} made an estimate
of the loss rate of particles,
$\dot{n}/n\approx111\left(na^{3}\right)^{2}\bar{\epsilon}/\hbar$
where $\bar{\epsilon}$ is the average kinetic energy of atoms.
Taking $\bar{\epsilon}\approx10\mu K$, a realistic energy scale in
current experiments, we estimate the life time of atoms in the
trap to be $\:10^{-6}s,\:10^{-3}s,\:10^{-2}s$ for
$g=2,\:0.4,\:0.2$ respectively, when the number of particles is
so large that the first node of the super-shell is reached. Hence
the life time is longer when $g$ (or $a$) is smaller, reflecting
that the loss rate is proportional to $a^{6}$. The temperature regime
of this super-shell structure is below $0.1 \mu K$. 

In conclusion, we have seen that the shell structure of 
fermions with weak, repulsive interactions
in a harmonic trap shows a pronounced beating
pattern, with the single shell positions changing by half a period
length between the different beat nodes. A Fourier analysis of the
oscillating shell-correction part of the Hartree-Fock energy shows
clear peaks at two slightly different frequencies. This is
interpreted semiclassically by the interference of the shortest
periodic orbits generated by the breaking of the U(3) symmetry of
the non-interacting HO system, which are the families of diameter
and circle orbits, through a uniform trace formula given fully in
\cite{semicl} and, in the perturbative limit, in Eq.\ (\ref{tfpert}).

For very weak  interactions, the splittings of the highly 
degenerate HO levels have earlier been calculated perturbatively 
within the WKB approximation~\cite{heiselberg}.
However, the perturbative results do not apply for
the interaction strengths 
where super-shell structure appears visible.

We acknowledge discussions with A. Bulgac, S. Creagh, S. Keppeler,
B. Mottelson and C. Pethick. This work was financially supported by
the Swedish Foundation for Strategic Research and the Swedish Research
Council. One of us (M.B.) acknowledges the warm hospitality at
the LTH during several research visits.

\end{document}